\documentclass[reprint,aip,apl,showpacs,superscriptaddress]{revtex4-1}
\usepackage{mathrsfs,bm,multirow,amsmath,amsfonts,amssymb,array,booktabs,float}
\usepackage{graphicx,times,graphics,color,epsfig}
\renewcommand{\thispagestyle}[1]{}

\begin{document}
\title{Impurity-limited quantum transport variability in magnetic tunnel junctions}
\author{Jianing Zhuang}
\affiliation{Department of Physics and the Center of Theoretical and Computational Physics, The University of Hong Kong, Pokfulam Road, Hong Kong SAR, China}
\affiliation{Institute of Textiles and Clothing, Hong Kong Polytechnic University, Kowloon, Hong Kong SAR, China}
\author{Yin Wang}
\email{yinwang@hku.hk}
\affiliation{Department of Physics and the Center of Theoretical and Computational Physics, The University of Hong Kong, Pokfulam Road, Hong Kong SAR, China}
\affiliation{The University of Hong Kong Shenzhen Institute of Research and Innovation, Shenzhen, Guangdong 518057, China}
\author{Yan Zhou}
\email{yanzhou@hku.hk}
\affiliation{Department of Physics and the Center of Theoretical and Computational Physics, The University of Hong Kong, Pokfulam Road, Hong Kong SAR, China}
\author{Jian Wang}
\affiliation{Department of Physics and the Center of Theoretical and Computational Physics, The University of Hong Kong, Pokfulam Road, Hong Kong SAR, China}
\affiliation{The University of Hong Kong Shenzhen Institute of Research and Innovation, Shenzhen, Guangdong 518057, China}
\author{Hong Guo}
\affiliation{Center for the Physics of Materials and Department of Physics, McGill University, Montreal, PQ, Canada, H3A 2T8}
\affiliation{Department of Physics and the Center of Theoretical and Computational Physics, The University of Hong Kong, Pokfulam Road, Hong Kong SAR, China}

\date{\today}
\begin{abstract}
We report an extensive first-principles investigation of impurity-induced device-to-device variability of spin-polarized quantum tunneling through Fe/MgO/Fe magnetic tunnel junctions (MTJ). In particular, we calculated the tunnel magnetoresistance ratio (TMR) and the average values and variances of the currents and spin transfer torque (STT) of an interfacially doped Fe/MgO/Fe MTJ. Further, we predicted that N-doped MgO can improve the performance of a doped Fe/MgO/Fe MTJ. Our first-principles calculations of the fluctuations of the on/off currents and STT provide vital information for future predictions of the long-term reliability of spintronic devices, which is imperative for high-volume production.
\end{abstract}

\maketitle

\emph{NOTE:} This article has been accepted by \emph{Frontiers of Physics} in a revised form.

\emph{Introduction.}
Device-to-device variability is an important figure-of-merit characterizing the robustness and fabrication reliability of electronic devices.\cite{ITRS} The variability arises from the inevitable atomistic disorder in all realistic systems.\cite{ITRS, Asenov, Koenraad} When the device size approaches tens of nanometers, the variability becomes significant, because each individual disorder configuration yields slightly different transport properties.\cite{Georgiev, Ohno} As circuit design becomes difficult if every device behaves differently, determining the source of this variability, along with its effect and magnitude, is becoming an increasingly important problem in nanoelectronic device physics. For example, a 10~nm device channel is only approximately 100 atoms long, even a single impurity or misplaced atom induces large property fluctuations, because such an impurity/atom yields 1\% disorder per structure width. In traditional complementary metal-oxide semiconductors (CMOS), device-to-device variability appears in many transport properties, including the conductance, the on and off currents, the threshold voltage, and the subthreshold swing.

While this variability is severe for traditional CMOS, it is expected to be even more problematic for spintronic devices, such as magnetic tunnel junctions (MTJs), which are the basic device elements of magnetic random access memory (MRAM).\cite{MRAM} There are several factors behind this expectation. First, the device physics of MTJs and MRAM is based on quantum tunneling of spin-polarized charges through very thin barriers, i.e., as thin as five layers of an insulator such as MgO.\cite{Parkin, Yuasa, Waldron} Clearly, for such thin layers, a misplaced atom or impurity will induce considerable disorder. As MTJs necessarily incorporate different materials, such as the insulating oxide barriers and magnetic metal contacts, it is difficult to construct an MTJ that is completely disorder-free, because of the mismatching at the material interfaces. Second, an important device merit of an MTJ is the tunnel magnetoresistance ratio (TMR), which has a sensitive dependence on the degree of spin polarization of the tunnel current. It has been shown that a small degree of interface disorder can be extremely detrimental to the TMR value.\cite{Ke} Third, the existence of magnetic impurities can severely affect the spin coherence, which is crucial for spintronics phenomena, and this effect also induces variability in the spin coherent transport. Given the importance of MTJ structures for many potential applications, such as MRAMs, ultra-sensitive magnetic field sensors, and spin torque nano-oscillators,\cite{MTJ, MRAM, STT, TMR} it is important to develop an understanding of the disorder-limited device-to-device variability of MTJs, which is the target of this work.

In particular, we report an extensive first-principles investigation of the impurity-limited device-to-device variability of spin-polarized quantum tunneling through an Fe/MgO/Fe MTJ. This system is chosen because it has the largest TMR among all MTJs investigated in the literature, and because of its widespread use in practical device applications. Our calculation is based on a recently developed first-principles approach (see below). We focus on both the average values and, more importantly, the variances of the TMR and spin transfer torque (STT) device properties. For the disorder, we focus on interfacial impurities. The predicted fluctuations in the TMR and STT provide quantitatively vital information regarding the variability magnitude versus the disorder strength, which is important in practical applications for which high-volume MTJ production is required.

\emph{Method.} Calculating the device-to-device variability from atomistic first-principles is, in fact, a very challenging problem.\cite{Zhuang} Typically, brute-force calculation of many disorder configurations for a given impurity concentration $x$ is required before the variance of a particular physical quantity can be obtained through statistical averaging. For quantum coherent transport, which is non-self-averaging, the disorder configuration ensemble can be very large. Furthermore, for the fully self-consistent atomistic modeling conducted in this work, such brute-force calculation is extremely inefficient, if not impossible, to perform from a computational perspective. Therefore, in this work, we adopt a recently developed non-equilibrium coherent potential approximation (NECPA) theory,\cite{NECPA} which \emph{analytically} performs the statistical configuration averaging that yields set expressions for the averaged transmission coefficient $T$ and the average of its square, in terms of various nonequilibrium Green's functions (NEGF). Using these expressions,\cite{Zhu,NECPA} first-principles numerical calculation needs to be done just once in order to obtain the desired transport variability.

Very briefly, our calculation proceeds as follows. The charge current $I$ is calculated from the transmission coefficient $T$ ($e=\hbar=1$), using\cite{Datta}
\begin{equation}
\label{eq1}
I\ =\ \int\frac{dE}{2\pi}T(E)[f_L(E)-f_R(E)],
\end{equation}
where $E$ is the electron energy and $f_L(E)$ and $f_R(E)$ are the Fermi functions of the left and right electrodes, respectively. The transmission coefficient $T$ is determined by the Green's functions, such that	
\begin{equation}
\label{trans1}
T(E)\ =\ \mathrm{Tr}[G^r(E)\Gamma_L(E)G^a(E)\Gamma_R(E)],
\end{equation}
where $G^{r,a}$ are the retarded and advanced Green's functions of the device scattering region, respectively, and $\Gamma_{L,R}$ are the line-width functions of the left and right electrodes, respectively. In the presence of disorder, the $G^{r,a}$ depend on the particular disorder configuration; hence, $T(E)$ also depends on this configuration. As a result, ensemble averaging of the physical quantity is required. We use $\langle{\cdots}\rangle$ to denote the ensemble average of the disorder configurations and, thus, the averaged current is expressed as
\begin{equation}
\langle I\rangle=\int\frac{dE}{2\pi}\langle{T(E)}\rangle[f_L(E)-f_R(E)],
\end{equation}
while the current fluctuation is\cite{Zhu}
\begin{eqnarray}\label{deltaI}
\delta I \equiv \sqrt{\langle {I^2}\rangle-\langle I\rangle^2}\approx \int\frac{dE}{2\pi}\delta{T(E)}[f_L(E)-f_R(E)],
\end{eqnarray}
with the transmission fluctuation being
\begin{equation}
\label{dT}
\delta T \equiv \sqrt{\langle{T^2}\rangle-\langle T\rangle^2}.
\end{equation}
Note that $\delta T$ measures the transport variability due to the disorder randomness; this is the central quantity of this work.

Calculating $\delta T$ requires $\langle{T(E)}\rangle$ and $\langle{T^2(E)}\rangle$. From Eq.~(\ref{trans1}), determining $\langle T\rangle$ requires evaluation of the correlation of \emph{two} Green's functions $\langle {G\cdot G}\rangle$; however, the calculation of $\langle{T^2(E)}\rangle$ necessitates the evaluation of the correlation of \emph{four} Green's functions, i.e., $\langle {G\cdot G\cdot G\cdot G}\rangle$. In Ref.~\onlinecite{Zhu}, these correlations are derived based on the NECPA theory,\cite{NECPA} in terms of the averaged single-particle Green's functions and various vertex correction terms.\cite{NVC} In realistic devices, considerable effort is devoted to control the disorder concentration $x$ to be very small (e.g., 1\% or less); thus, a low-concentration approximation is further derived in Ref.~\cite{Zhu}, so as to simplify the expressions of the correlation functions $\langle {G\cdot G}\rangle$ and $\langle {G\cdot G\cdot G\cdot G}\rangle$, namely to express $\delta T$ to the first order in the disorder concentration $x$. These theoretical derivations are quite tedious and will not be reproduced here. Instead, we refer interested readers to the original works\cite{Zhu,NECPA, Eric} for the relevant derivations.

In our numerical calculations, we use the quantum transport package Nanodsim,\cite{NECPA, IEEEreview, Eric} which implements various levels of the NECPA theory, including the low-concentration approximation of $\delta T$. In this method, density functional theory (DFT) is carried out within the NEGF, such that the Hamiltonian of the open device structure is calculated self-consistently without phenomenological parameters.\cite{NEGFDFT} The disorder scattering and impurity-limited device-to-device variability are treated at the level of local concentration approximation of the NECPA, which has been shown to yield very accurate results\cite{Zhu} for the concentrations considered herein.

\emph{Device model and physical quantities.} Without loss of generality, we consider the most popular MTJ, which is comprised of Fe/MgO/Fe(001). Ref.~\onlinecite{TMR} has elegantly explained the microscopic physics of the operation of an ideal, disorder-free Fe/MgO/Fe junction. That is, by symmetry the minority-spin $d$-states having transverse momentum $k_{\|}\neq(0,0)$ in Fe are filtered by MgO, because they cannot couple to the slowly decaying MgO $\Delta_1$ band at $k_{\|}=(0,0)$. In addition, the majority-spin channel in the left Fe cannot tunnel through the junction when the right Fe is in the antiparallel state. The overall result is a very small current for the antiparallel state and a large spin-polarized current for the parallel state, inducing an extremely large TMR in the device. Clearly, the physics of this microscopic device is affected significantly by the occurrence of disorder; for example, Ref.~\onlinecite{Ke} has shown that a small number of interfacial O vacancies diminishes the TMR value considerably. In the remainder of this paper, we investigate the device-to-device variability resulting from such disorder effects, as quantified by $\delta T$.

\begin{figure}[tbp]\vspace{0.2cm}
\includegraphics[width=0.5\textwidth]{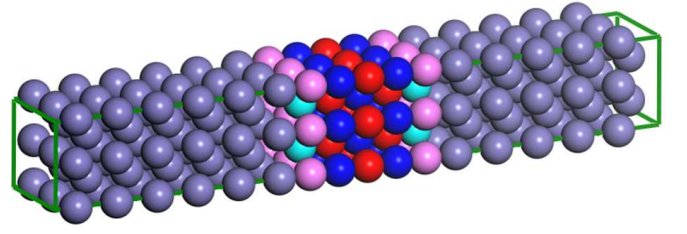}
\caption{(Color online). Atomic structure of Fe/MgO/Fe MTJ with 5-layer MgO barrier. The gray, dark and light-blue, and red and pink spheres indicate Fe, Mg, and O, respectively. Because of the atomic disorder, some sites in the MgO region near both Fe/MgO interfaces are doped with vacancies, N, or Al, as indicated by the light-blue and pink spheres. The MTJ is periodically extended in the transverse direction.}\label{Struct}
\end{figure}

Fig.~\ref{Struct} illustrates the atomic structure of an Fe/MgO/Fe(001) MTJ, which is a two-probe transport junction. The MTJ consists of an atomically thin insulating barrier (MgO) sandwiched between two ferromagnetic leads (Fe). The magnetic moments of the two leads can be in a parallel (PC) or antiparallel (APC) configuration. The impurities located near the Fe/MgO interfaces contribute significantly to the variability. Thus, we model these interfacial impurities by introducing an impurity-atom concentration (again represented by $x$) in the MgO barrier near the Fe/MgO interfaces; these impurities include O vacancies on the O sites or N (Al) impurity atoms on the O (Mg) sites.

The TMR of the MTJ is defined by both the on-state current through the PC ($I_{\uparrow\uparrow}$) and the off-state current through the APC ($I_{\uparrow\downarrow}$), as the $TMR=(I_{\uparrow\uparrow}-I_{\uparrow\downarrow})/I_{\uparrow\downarrow}$. A large TMR value indicates that the on- and off-state currents are well separated, i.e., $I_{\uparrow\uparrow}\gg I_{\uparrow\downarrow}$.
Another important quantity is the STT.\cite{STT, STT2} Note that a larger STT for a given input current density implies higher-efficiency switching between the PC and APC as a result of the spin-polarized current. Using $I_{\uparrow\uparrow}^\uparrow$ and $I_{\uparrow\uparrow}^\downarrow$ to denote the spin-up and -down currents of the PC, respectively, and $I_{\uparrow\downarrow}^\uparrow$ and $I_{\uparrow\downarrow}^\downarrow$ for their APC counterparts, the transverse (in-plane) STT can be obtained from \cite{STT_eq, Jia}
\begin{equation}\label{STT_formul}
\bold{T}_{||}(\theta)=\frac{\hbar}{2e}\frac{(I_{\uparrow\downarrow}^\uparrow-I_{\uparrow\downarrow}^\downarrow)-(I_{\uparrow\uparrow}^\uparrow-I_{\uparrow\uparrow}^\downarrow)}{2}\bold{M_2}
\times(\bold{M_1}\times\bold{M_2}),
\end{equation}
in which $\bold{M_1}$ and $\bold{M_2}$ represent the magnetization of the left and right leads, respectively, and $\theta$ is the relative angle between them. For a specific lead material and $\theta$, the angular component $\bold{M_2}\times(\bold{M_1}\times\bold{M_2})$ provides a constant vector in $\bold{T}_{||}(\theta)$. In our calculations, we focus on the zero-bias limit (linear regime), where all the currents are replaced by the corresponding $T$ [via Eq.~(\ref{eq1})].

Eq.~(\ref{dT}) is employed in order to obtain a measure of the device-to-device variability. For spin-polarized quantum transport, the PC transmission
$T_{\uparrow\uparrow}=T_{\uparrow\uparrow}^\uparrow + T_{\uparrow\uparrow}^\downarrow$. That is, $T_{\uparrow\uparrow}$ is due to the two spin transport channels indicated by the superscripts. While the configuration average is simple for the transmission itself, with $\langle T_{\uparrow\uparrow}\rangle =\langle T_{\uparrow\uparrow}^\uparrow\rangle + \langle T_{\uparrow\uparrow}^\downarrow\rangle$, the average over the transmission square $\langle T_{\uparrow\uparrow}^2 \rangle$ induces a cross term $\langle T_{\uparrow\uparrow}^\uparrow \times T_{\uparrow\uparrow}^\downarrow\rangle$, which may not be equal to $\langle T_{\uparrow\uparrow}^\uparrow\rangle \times \langle T_{\uparrow\uparrow}^\downarrow\rangle$ if magnetic impurities are present in the tunnel barrier to induce spin-flip scattering. However, in this work, since we investigate non-magnetic impurities and furthermore we neglect the very small magnetic proximity effect on the impurity sites - note that the impurity concentration $x$ is very small, $\langle T_{\uparrow\uparrow}^\uparrow \times T_{\uparrow\uparrow}^\downarrow\rangle= \langle T_{\uparrow\uparrow}^\uparrow\rangle \times \langle T_{\uparrow\uparrow}^\downarrow\rangle$.

The NEGF-DFT calculations are performed within the linear muffin-tin orbital (LMTO) scheme and the atomic sphere approximation (ASA) framework,\cite{LMTO} implemented in the Nanodsim quantum transport package.\cite{NECPA, IEEEreview, Eric} A $40 \times 40$ k-mesh was used to sample the transverse Brillouin zone during the self-consistency in order to converge the density matrix, and a $400 \times 400 $ k-mesh was used to calculate the spin-resolved transmission coefficients $T$ and their fluctuations
$\delta T$, as described above. The MTJ atomic structure was relaxed using the DFT total energy method, following the standard approach described in the literature.\cite{Ke}

\begin{figure}[tbp]\vspace{0.2cm}
\includegraphics[width=0.5\textwidth]{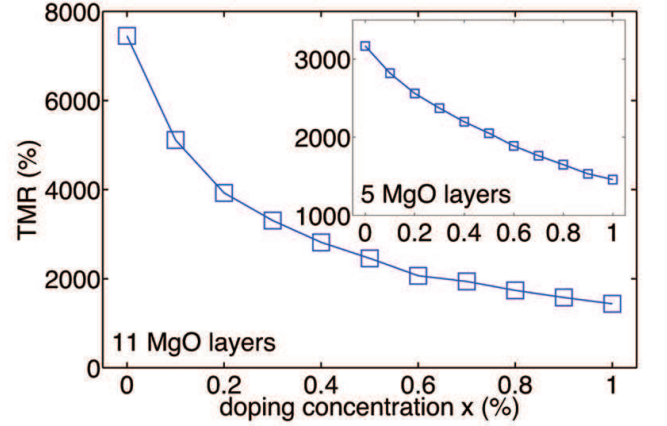}
\caption{Average TMR values vs. O-vacancy concentration $x$ for MTJs having 5 and 11 MgO layers.}\label{TMR}
\end{figure}

\emph{Results.}
Fig.~\ref{TMR} shows the calculated TMR values for MTJs consisting of 5 and 11 MgO layers with different concentrations $x$ of O vacancies at both Fe/MgO interfaces. The TMR values decrease dramatically for both devices with increasing $x$, which is in agreement with previous calculations.\cite{Ke} Note that, for O-vacancy $x=1\%$, the TMR values of both MTJs remain high (approximately 1500\%), indicating that the on- and off-state are well separated and the digital system can function.

\begin{figure}[tbp]\vspace{0.2cm}
\includegraphics[width=0.5\textwidth]{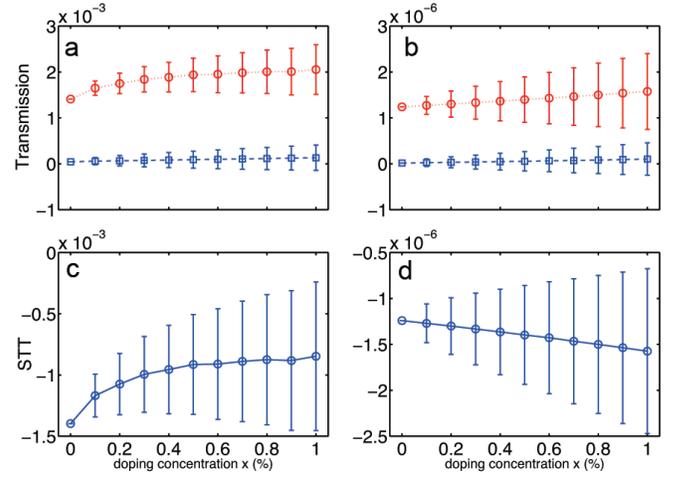}
\caption{(Color online) (a, b) On- (red dotted lines) and off-state (blue dashed lines) currents for MTJs with 5 and 11 MgO layers, respectively. (c, d) STT for MTJs with 5 and 11 MgO layers, respectively. The error bars show the variances quantifying the device-to-device variability.}\label{current-STT}
\end{figure}

Figs.~\ref{current-STT}(a) and (b) show the average values and variance of the on- and off-state currents for MTJs with 5 and 11 MgO layers, respectively. With increasing $x$, the current fluctuation increases dramatically. For $x=1\%$, the variability value is comparable to the average current value. Larger fluctuation of the on- and off-state currents induces ambiguity in the two states, although the ``average TMR" value remains large [Fig.~\ref{TMR}]. Note that our fluctuation calculations provide a quantitative description of the half-width of the current distribution; therefore, the negative values presented for the off-state current fluctuations do not indicate negative current values.

Figs. \ref{current-STT}(c) and (d) show the transverse STT and its fluctuation for MTJs having 5 and 11 MgO layers. For the entire calculated doping-concentration range, $0<x<1\%$, the average STT values remain almost unchanged, since the spin-up on-state current dominates the STT value. However, with increasing $x$, the STT fluctuations increase dramatically and become comparable to the average value at $x=1\%$.

\begin{table}[htbp]\footnotesize
\begin{tabular}{|l|l|l|l|l|l|l|l|l|}
\hline
System      & TMR($\%$) & $\langle T_{on}\rangle$ & $\langle T_{off}\rangle$ & $\langle STT \rangle$ & $\delta T_{on}$ & $\delta T_{off}$ & $\delta STT$ \\\hline
Clean        &  7466.6   & 1.24e-6                 & 1.64e-8                  & -1.24e-6              & 0               & 0                & 0      \\\hline
Vac            &  2458.0   & 1.40e-6                & 5.46e-8                  & -1.40e-6              & 4.96e-7         & 2.10e-7          & 5.38e-7      \\\hline
Al              &  4687.1   & 1.68e-6                 & 3.52e-8                  & -1.68e-6              & 6.80e-7         & 3.00e-8          & 6.80e-7      \\\hline
N               &  7662.0   & 1.01e-6                 & 1.30e-8                  & -1.01e-6              & 7.49e-9         & 7.81e-10          & 7.53e-9      \\\hline
\end{tabular}
\caption{TMR, on- and off-state transmission coefficients $T_{on}$ and $ T_{off}$, respectively, STT, and their fluctuations (indicated by $\delta$) for MTJ with 11 MgO layers. $x=0.5\%$ for both Fe/MgO interfaces for the vacancy (Vac)-, N-, and Al-doped systems.}
\label{eledope}
\end{table}

Having known that the O vacancies at the anion sites will lead toinduce considerable fluctuation of the currents and STT values, we also investigated the doping effects at the cation sites. Considering the fact that Al$_2$O$_3$ is a widely used barrier material for MTJs, Al atoms at $x=0.5\%$ were used to replace the Mg atoms at both interfaces for an MTJ with 11 MgO layers. The calculated results are listed in TABLE~\ref{eledope}. It is apparent that, although the TMR value of the Al-doped MTJ (4867.1\%) is significantly higher than that of the vacancy-doped MTJ (2458.0\%), the fluctuation values of both systems have almost the same order of magnitude ($10^{-7}$).

The filling of O vacancies with N at the anion sites can maintain a high TMR for an Fe/MgO/Fe MTJ;\cite{Ke, Liu} thus, it is worth investigating whether the presence of N-doped MgO at the anion sites will reduce the current fluctuation values. We performed calculations with $x=0.5\%$ interfacially doped N. Again, the results are listed in TABLE~\ref{eledope}. The fluctuation values of both currents and STT are reduced by two or three orders compared to the MTJ with O vacancies, indicating that N impurities are beneficial to the Fe/MgO/Fe MTJ device performance. These results are reasonable, because N has a similar electron configuration and atomic radius to O, and reduces the disorder scattering at the interfaces.

In summary, we used the LMTO-NEGF-DFT approach to predict the device-to-device variability of an Fe/MgO/Fe(001) MTJ with different doping concentrations and dopants. Various quantities related to the device-to-device variability were studied, including the on- and off-state currents, the STT, and the TMR. The results of these calculations show that, although the TMR values remain at a reasonably high value, the impurity-induced device-to-device variability can strongly affect the device performance. In particular, we demonstrated that N-doped MgO can improve the performance of an Fe/MgO/Fe MTJ compared to the same MTJ with vacancies at the interfaces. Our LMTO-NEGF-DFT approach for predicting the device-to-device variability will provide vital information for device physics and engineering in future research.

Yin Wang is grateful to Drs. Yu Zhu, Lei Liu, Lei Zhang, and Dongping Liu for useful discussion. This work is supported by the University Grant Council (Contract No. AoE/P-04/08) of the Government of HKSAR (YW, JW), the HKU Seed Funding Programme for Basis Research (YW), NSERC and IRAP of Canada (HG), and the NSFC (11404273). We thank CLUMEQ and Compute-Canada for the provision of computational resources.


\vspace{0.5cm}

\begin{thebibliography}{99}


\bibitem{ITRS}
International technology roadmap for semiconductors, http://public.itrs.net/.
\bibitem{Asenov}
A. Asenov, IEEE Trans. Electron Devices {\bf 45}, 2505 (1998).
\bibitem{Koenraad}
P. M. Koenraad and M. E. Flatt\'e, Nature Materials {\bf 10}, 91 (2011).
\bibitem{Ohno}
H. Ohno, Science {\bf 281}, 951 (1998).
\bibitem{Georgiev}
V. P. Georgiev, E. A. Towie, and A. Asenov, IEEE Trans. Electron Deveces {\bf 60}, 965 (2013).
\bibitem{MRAM}
S. Ikeda, J. Hayakawa, Y. M. Lee, F. Matsukura, Y. Ohno, T. Hanyu, and H. Ohno, IEEE Trans. Electron Devices {\bf 54}, 991 (2007); J. A. Katine and E. E. Fullerton, J. Magn. Magn. Mater. {\bf 320}, 1217 (2008).
\bibitem{Parkin}
S.~S.~P.~Parkin, C.~Kaiser, A.~Panchula, P.~M.~Rice, B.~Hughes, M.~Samant, and S.-H.~Yang, Nat.~Mater. {\bf 3}, 862 (2004).
\bibitem{Yuasa}
S.~Yuasa, T.~Nagahama, A.~Fukushima, Y.~Suzuki, and K.~Ando, Nat.~Mater. {\bf 3}, 868 (2004).
\bibitem{Waldron}
D.~Waldron, P.~Haney, B.~Larade, A.~MacDonald, and H.~Guo, Phys.~Rev.~Lett. {\bf 96}, 166804 (2006).
\bibitem{Ke}
Y. Ke, K. Xia, and H. Guo, Phys. Rev. Lett. {\bf 105}, 236801 (2010).
\bibitem{MTJ}
S. S. P. Parkin, C. Kaiser, A. Panchula, P. M. Rice, B. Hughes, M. Samant, and S.-H. Yang, Nature Mater. {\bf 3}, 862 (2004); S. Yuasa, T. Nagahama, A. Fukushima, Y. Suzuki, and K. Ando, Nature Mater. {\bf 3}, 868 (2004).
\bibitem{STT}
J. C. Slonczewski, J. Magn. Magn. Master. {\bf 159}, L1 (1996); L. Berger, Phys. Rev. B {\bf 54}, 9353 (1996).
\bibitem{TMR}
W. H. Butler, X.-G. Zhang, T. C. Schulthess, and J. M. Maclaren, Phys. Rev. B {\bf 63}, 054416 (2001).
\bibitem{Zhuang}
J. Zhuang and J. Wang, J. Appl. Phys. {\bf 114}, 063708 (2013).
\bibitem{NECPA}
Y. Zhu, L. Liu, and H. Guo, Phys. Rev. B {\bf 88}, 205415 (2013).
\bibitem{Zhu}
Y. Zhu, L. Liu, and H. Guo, Phys. Rev. B {\bf 88}, 085420 (2013).
\bibitem{Datta}
S. Datta, \textit{Electronic Transport in Mesoscopic System} (Cambridge University Press, Cambridge, England, 1995).
\bibitem{NVC}
Y. Ke, K. Xia, and H. Guo, Phys. Rev. Lett. {\bf 100}, 166805 (2008).
\bibitem{IEEEreview}
J. Maassen, M. Harb, V. Michaud-Rioux, Y. Zhu, and H. Guo, Proc. IEEE {\bf 101}, 518 (2013).
\bibitem{Eric}
Y.~Zhu and L.~Liu, \emph{Atomistic Simulation of Quantum Transport in Nanoelectronic Devices}, (Workd Scientific, 2016).
\bibitem{NEGFDFT}
J. Taylor, H. Guo and J. Wang, Phys. Rev. B {\bf 63}, 121104(R) (2001); {\bf 63}, 245407 (2001).
\bibitem{STT2}
S. I. Kiselev, J. C. Sankey, I. N. Krivorotov, N. C. Emley, M. Rinkoski, C. Perez, R. A. Buhrman, and D. C. Ralph, Phys. Rev. Lett. {\bf 93}, 036601 (2004); D. C. Ralph and M. D. Stiles, J. Magn. Magn. Mater. {\bf 320}, 1190 (2008); P. M. Haney, R. A. Duine, A. S. Nunez, and A. H. MacDonald, \textit{ibid.} {\bf 320}, 1300 (2008).
\bibitem{STT_eq}
I. Theodonis, N. Kioussis, A. Kalitsov, M. Chshiev, and W. H. Butler, Phys. Rev. Lett. {\bf 97}, 237205 (2006).
\bibitem{Jia}
X. Jia, K. Xia, Y. Ke, and H. Guo, Phys. Rev. B {\bf 84}, 014401 (2011).
\bibitem{LMTO}
I. Turek, V. Drchal, J. Kudrnovsk\'y, M. \v{S}ob, and P. Weinberger, \textit{Electronic Structure of the Disordered Alloys, Surfaces, and Interfaces} (Kluwer, Boston, 1977); J. Kudrnovsk\'y, V. Drchal, and J. Masek, Phys. Rev. B {\bf 35}, 2487 (1987).
\bibitem{Liu}
D. Liu, X. Han, and H. Guo, Phys. Rev. B {\bf 85}, 245436 (2012).


\end{thebibliography}
\end{document}